\documentclass[aps,prl,preprintnumbers,twocolumn,groupedaddress,nofootinbib]{revtex4}

\usepackage[dvips]{graphicx}
\usepackage{color}
\usepackage{amsmath,amssymb,slashed}
\usepackage{hyperref}

\usepackage{changepage}

\newif\ifdraft
\drafttrue
\newif\ifpreprint
\preprinttrue
\def\spa#1.#2{\left\langle#1\,#2\right\rangle}
\def\spb#1.#2{\left[#1\,#2\right]}

\makeatletter
\providecommand*{\shuffle}{%
  \mathbin{\mathpalette\shuffle@{}}%
}
\newcommand*{\shuffle@}[2]{%
  \sbox0{$#1\vcenter{}$}%
  \kern .15\ht0 % side bearing
  \rlap{\vrule height .25\ht0 depth 0pt width 2.5\ht0}%
  \raise.1\ht0\hbox to 2.5\ht0{%
    \vrule height 1.75\ht0 depth -.1\ht0 width .17\ht0 %
    \hfill
    \vrule height 1.75\ht0 depth -.1\ht0 width .17\ht0 %
    \hfill
    \vrule height 1.75\ht0 depth -.1\ht0 width .17\ht0 %
  }%
  \kern .15\ht0 % side bearing
}
\makeatother

\makeatletter
\providecommand*{\boldshuffle}{%
  \mathbin{\mathpalette\boldshuffle@{}}%
}
\newcommand*{\boldshuffle@}[2]{%
  \sbox0{$#1\vcenter{}$}%
  \kern .15\ht0 % side bearing
  \rlap{\vrule height .25\ht0 depth 0.25pt width 2.5\ht0}%
  \raise.1\ht0\hbox to 2.5\ht0{%
    \vrule height 1.75\ht0 depth -.1\ht0 width .31\ht0 %
    \hfill
    \vrule height 1.75\ht0 depth -.1\ht0 width .31\ht0 %
    \hfill
    \vrule height 1.75\ht0 depth -.1\ht0 width .31\ht0 %
  }%
  \kern .15\ht0 % side bearing
}
\makeatother

\def\beq{\begin{equation}}
\def\eeq{\end{equation}}

\newcommand{\mot}{\mathfrak m}
\newcommand{\dR}{\mathfrak{dr}}
\newcommand{\MM}{\mathbb M}

\newcommand{\GG}{\mathbb G}
\newcommand{\FFF}{\mathbb F}

\newcommand{\eq}{\begin{equation}}
\newcommand{\eqe}{\end{equation}}
\newcommand{\eqa}{\begin{eqnarray}}
\newcommand{\eqae}{\end{eqnarray}}

\newcommand{\bea}{\begin{eqnarray}}
\newcommand{\eea}{\end{eqnarray}}
\newcommand{\dd}{\mathrm{d}}

\newcommand{\bma}{\begin{matrix}}
\newcommand{\ema}{\cr\end{matrix}}

\newcommand{\CC}{\mathbb C}
\newcommand{\NN}{\mathbb N}

\newcommand{\QQ}{\mathbb Q}

\def\NN{{\mathbb N}}
\def\CC{{\mathbb C}}

\def\QQ{{\mathbb Q}}
\def\GG{{\mathbb G}}

\def\mylength{-0.35cm}
\def\myotherlength{-0.25cm}

\newbox\charbox
\newbox\slabox
\def\s#1{{     % Feynman slash
        \setbox\charbox=\hbox{$#1$}
        \setbox\slabox=\hbox{$/$}
        \dimen\charbox=\ht\slabox
        \advance\dimen\charbox by -\dp\slabox
        \advance\dimen\charbox by -\ht\charbox
        \advance\dimen\charbox by \dp\charbox
        \divide\dimen\charbox by 2
        \raise-\dimen\charbox\hbox to \wd\charbox{\hss/\hss}
        \llap{$#1$}
}}

\begin{document}

\preprint{UUITP--36/23}

\title{Motivic coaction and single-valued map of polylogarithms from zeta generators}

\author{Hadleigh Frost$^a$, Martijn Hidding$^{b,c}$, Deepak Kamlesh$^a$, Carlos Rodriguez$^{b,}$\footnote{Corresponding author, carlos.rodriguez@physics.uu.se}, Oliver Schlotterer$^{b}$ and Bram Verbeek$^b$}
\affiliation{$^a$ Mathematical Institute, University of Oxford, OX2 6GG Oxford, United Kingdom}
\affiliation{$^b$ Department of Physics and Astronomy, Uppsala University, Box 516, 75120 Uppsala, Sweden}
\affiliation{$^c$ Institute for Theoretical Physics, ETH Zurich, %
8093 Z\"urich, Switzerland}

\begin{abstract}
We introduce a new Lie-algebraic approach to explicitly construct the motivic coaction 
and single-valued map of multiple polylogarithms in any number of variables. In both cases, 
the appearance of multiple zeta values is controlled by conjugating generating series 
of polylogarithms with Lie-algebra generators associated with odd zeta values. Our 
reformulation of earlier constructions of coactions and single-valued polylogarithms 
preserves choices of fibration bases, exposes the correlation between multiple zeta 
values of different depths and paves the way for generalizations beyond genus zero.
\end{abstract}

\maketitle

%%%%%%%%%%%%%%%%%%%%%%%%%%%%%%%%%%%%%%%%%%%%%%%%%%%%%%%%%
%%%%%%%%%%%%%%%%%%%%%%%%%%%%%%%%%%%%%%%%%%%%%%%%%%%%%%%%%
\section{Introduction}
%%%%%%%%%%%%%%%%%%%%%%%%%%%%%%%%%%%%%%%%%%%%%%%%%%%%%%%%%
%%%%%%%%%%%%%%%%%%%%%%%%%%%%%%%%%%%%%%%%%%%%%%%%%%%%%%%%%
\vspace{\myotherlength}

The advent of multiple polylogarithms (MPLs) \cite{GONCHAROV1995197, Goncharov:1998kja, Remiddi:1999ew, Goncharov:2001iea, Vollinga:2004sn} led to a wealth of new perturbative
results and structural insights in quantum field theories \cite{Duhr:2014woa, Henn:2014qga, Bourjaily:2022bwx, Abreu:2022mfk, Weinzierl:2022}, string theory \cite{Berkovits:2022ivl, Mafra:2022wml} and 
numerous other
areas of theoretical physics where iterated integrals of rational functions are encountered.
Apart from systematizing the integration over marked points on the sphere, MPLs
enjoy Hopf-algebra structures including the motivic coaction and the single-valued map which 
exposed surprising unifying phenomena in the recent amplitudes literature.

The motivic coaction \cite{Goncharov:2001iea, Goncharov:2005sla, BrownTate, Duhr:2012fh, Brown2014MotivicPA} mapping MPLs to tensor products of similar iterated integrals
turns out to stabilize and connect perturbative quantities in a variety of quantum field theories
and string theories \cite{Schnetz:2013hqa, Brown:2015fyf}. Examples include periods in $\phi^3$ and $\phi^4$ theory \cite{Panzer:2016snt, Borinsky:2022lds}, the anomalous magnetic moment 
of the electron \cite{Schnetz:2017bko}, ${\cal N}{=}4$ super-Yang--Mills \cite{Caron-Huot:2019bsq}, various families of Feynman integrals \cite{Abreu:2014cla, Abreu:2015zaa, Abreu:2017enx, Abreu:2017mtm, Tapuskovic:2019cpr, Abreu:2019xep, Gurdogan:2020ppd}, 
the disk integrals in string tree-level amplitudes \cite{Schlotterer:2012ny, Drummond:2013vz} and
the antipodal duality between amplitudes and form factors \cite{Dixon:2021tdw, Dixon:2022xqh, Dixon:2023kop}.
In this broad spectrum of cases, the motivic 
coaction of MPLs in physical quantities hints at novel types of symmetries under the so-called cosmic Galois group \cite{Cartier:1988, Brown:2015fyf}. 
Such Galois symmetries extract an infinite pool of perturbative information from amplitude contributions at low loop orders.

The single-valued map \cite{Schnetz:2013hqa, Brown:2013gia, Brown:2018omk} in turn solves the problem of eliminating the monodromies of
MPLs by systematically adding complex conjugate MPLs and multiple zeta values (MZVs),
while preserving holomorphic differential equations. The resulting single-valued MPLs \cite{svpolylog}
and their associated single-valued MZVs \cite{Schnetz:2013hqa, Brown:2013gia} became key ingredients for
studying the multi-Regge regime of ${\cal N}=4$ super-Yang--Mills \cite{Dixon:2012yy, DelDuca:2016lad, Broedel:2016kls, DelDuca:2019tur}, the high-energy limit of more general gauge theories \cite{DelDuca:2013lma, DelDuca:2017peo}
and numerous classes of Feynman
integrals \cite{Drummond:2012bg, Schnetz:2013hqa, BROWN2015478, Duhr:2023bku}. Moreover, the single-valued map is a number-theoretic bridge between 
open- and closed-string interactions in flat spacetime,
with a detailed understanding at tree level \cite{Schlotterer:2012ny, Stieberger:2013wea, Stieberger:2014hba, Schlotterer:2018abc, Vanhove:2018elu, Brown:2019wna} and first loop-level echoes in \cite{DHoker:2015wxz, Broedel:2018izr, Gerken:2018jrq, Zagier:2019eus, Gerken:2020xfv}. Finally, the single-valued map
has initiated bootstrap approaches to string amplitudes in AdS backgrounds~\cite{Alday:2022xwz, Alday:2023jdk, Alday:2023mvu, Fardelli:2023fyq}.

In this work, we provide a new construction of both the motivic coaction and
the single-valued map of MPLs in any number of variables. 
The main novelty is the use of Lie-algebra structures that efficiently reorganize the 
appearance of MZVs and lower-complexity MPLs. Apart from the
well-known braid operators for marked points on a sphere, our construction
makes essential use of Lie-algebra generators associated with odd Riemann zeta
values \cite{DG:2005, Brown:depth3}. While state-of-the-art formulations of both the motivic coaction \cite{Goncharov:2001iea, Goncharov:2005sla, Duhr:2012fh, Abreu:2017mtm} 
and single-valued MPLs \cite{svpolylog, DelDuca:2016lad, Broedel:2016kls} 
in principle cover cases in any number 
of variables and of arbitrary transcendental weight, the advantages of 
our alternative construction are
\vspace{-0.1cm}
\begin{itemize}
\item[(i)] The results are in fully simplified form with respect to relations among MPLs and MZVs over~$\mathbb Q$, 
%and 
i.e.\ are automatically cast into a fibration basis of~MPLs.
\vspace{-0.2cm}
\item[(ii)] The appearance of odd Riemann zeta values completely determines
the coefficients of MZVs beyond depth one and products
of arbitrary MZVs.
\vspace{-0.2cm}
\item[(iii)] The Lie-algebraic approach inspires generalizations beyond genus zero
where additional generators are associated to all moduli of the surface.
\end{itemize}
\vspace{-0.1cm}
While the first point (i) may be viewed as a computational benefit, the exposure of
correlations (ii) between MZVs of different depths is a structural virtue of
our approach to coaction formulas and single-valued periods. The last point (iii) already found an
explicit genus-one manifestation in section 4.1.2 of \cite{Dorigoni:2022npe}\footnote{In follow-up work \cite{Dorigoni:2023part2},
generating series of single-valued iterated Eisenstein integrals will be shown to
exhibit even closer parallels to the genus-zero results of this work once the
non-geometric parts $z_{2k+1}$ of the zeta generators in \cite{Dorigoni:2022npe} are completed
to generators $\sigma_{2k+1}$ including Tsunogai's derivations dual to holomorphic Eisenstein series \cite{Ihara:1990, IharaTakao:1993, Tsunogai, Pollack}.} 
for Brown's single-valued iterated Eisenstein integrals \cite{Brown:mmv, Brown:2017qwo, Brown:2017qwo2} which indicates that the series of 
zeta generators in this work are in fact a genus-agnostic building block. In this way, our results 
should unlock new perspectives on the appearance of MZVs and genus-zero MPLs in coaction
formulas for elliptic MPLs \cite{Broedel:2018iwv, Wilhelm:2022wow, Forum:2022lpz, Tapuskovic:2023xiu} 
and ultimately their generalizations to higher-genus surfaces or even higher-dimensional varieties.
Hence, the results of this work are hoped to serve as a stepping stone towards an all-loop
understanding of the role of the cosmic Galois group in quantum field theory and string theory,
as well as the interplay of open- and closed-string interactions in different backgrounds.

\vspace{\mylength}
%%%%%%%%%%%%%%%%%%%%%%%%%%%%%%%%%%%%%%%%%%%%%%%%%%%%%%%%%
%%%%%%%%%%%%%%%%%%%%%%%%%%%%%%%%%%%%%%%%%%%%%%%%%%%%%%%%%
\section{Review of Multiple Polylogarithms}
%%%%%%%%%%%%%%%%%%%%%%%%%%%%%%%%%%%%%%%%%%%%%%%%%%%%%%%%%
%%%%%%%%%%%%%%%%%%%%%%%%%%%%%%%%%%%%%%%%%%%%%%%%%%%%%%%%%
\vspace{\myotherlength}

We follow the standard conventions in \cite{Goncharov:2001iea} and define multiple polylogarithms (MPLs) recursively by
\beq
G(a_1,a_2,\ldots,a_w;z) = \int^z_0 \frac{ \dd t}{t-a_1} \, G(a_2,\ldots ,a_w;t)
\label{coact.01}
\eeq
with labels $a_1,\ldots,a_w\in \CC$, argument $z \in \CC$, (transcendental) weight~$w \in \NN$
and convention $G(\emptyset;z) = 1$.
Endpoint divergences are shuffle-regularized\footnote{Shuffle-regularization amounts
to imposing that the shuffle product $G(\vec{a};z)G(\vec{b};z)
= \sum_{\vec{c} \in \vec{a} \shuffle \vec{b}}G(\vec{c};z)$ (for ordered 
sets $\vec{a},\vec{b},\vec{c}$ of labels) universal to iterated
integrals extends to regularized values of formally divergent cases.} with the assignment
$G(0;z) = \log(z)$ and $G(z;z) = {-} \log(z)$ of regularized values at
weight one (see e.g.\ \cite{Panzer:2015ida}). Evaluation of MPLs with the restricted alphabet $a_i \in \{0,1\}$ at $z=1$ yields multiple zeta values (MZVs)
\begin{align}
&\zeta_{n_1,n_2,\ldots,n_r} = \sum^\infty_{0<k_1 <k_2<\ldots <k_r} 
k_1^{-n_1} k_2^{-n_2}\ldots k_r^{-n_r}
\label{coact.02}\\
&\quad = (-1)^r G(\underbrace{0,\ldots,0}_{n_r-1},1,\ldots,
\underbrace{0,\ldots,0}_{n_2-1},1,\underbrace{0,\ldots,0}_{n_1-1},1;1)
\notag
\end{align}
of depth $r$ and weight $n_1{+}\ldots{+}n_r$ with  $n_i \in \NN$ and $n_r {\geq} 2$. 

\medskip
\noindent
{\bf A. Motivic coaction:} Goncharov's and Brown's work \cite{Goncharov:2001iea, Goncharov:2005sla, Brown:2011ik, BrownTate}
gives an explicit formula for the motivic coaction of MPLs (\ref{coact.01})
at arbitrary $a_i, z \in \CC$.
The outcome of the coaction is conventionally
denoted by a tensor product, where the first and second entry
are mathematically understood as motivic periods and de Rham periods, respectively; see \cite{Goncharov:2005sla, BrownTate, Brown2014MotivicPA, Francislecture} 
for the deep algebraic-geometry background of these two types of periods. 
We will drop the tensor product and rely on  superscripts to distinguish between the motivic and the de Rham entry: throughout this work, we will write
$G^{\mot}$ instead of $ G^{\mot} \otimes 1$ and $G^{\dR}$ instead of $ 1\otimes G^{\dR}$
 (and similarly for $\zeta^{\mot},\zeta^{\dR}$).
De Rham periods
are only defined up to discontinuities such that powers of $i\pi$ are set to zero
in the second entry of the coaction, 
i.e.\ we have $\zeta_{2k}^{\dR}=0$.

A simple class of terms in the motivic coaction of MPLs is obtained from
deconcatenating its labels $a_i$
\begin{align}
\Delta G^{\mot}(a_1,a_2,\ldots,a_w;z)&= \sum_{j=0}^w  G^{\mot}(a_{j+1},\ldots,a_{w} ;z)
 \label{coact.06}  \\
 &\quad  \times G^{\dR}(a_1,\ldots,a_j;z) + \ldots
\notag
\end{align}
However, the coaction formula of \cite{Goncharov:2001iea, Goncharov:2005sla} typically 
yields numerous additional terms in the ellipsis which are more challenging to
state in closed form and involve at least one factor of $\zeta^{\dR}$ or $ G^{\dR}(\ldots;a_i)$ with $a_i \neq z$.
The simplest example,
% in the one-variable case, with $a_i \in \{0,1\}$, is
with $a_i \in \{0,1\}$, is
the last term of 
\small
\begin{align}
&\Delta G^{\mot}(0,0,1,1;z) = G^{\dR}(0,0,1,1;z)
+ G^{\mot}(1;z) G^{\dR}(0,0,1;z)
 \notag \\
&\quad\quad\quad\quad\quad
+G^{\mot}(1,1;z) G^{\dR}(0,0;z)
+ G^{\mot}(0,1,1;z) G^{\dR}(0;z) \notag \\
&\quad\quad\quad\quad\quad
+G^{\mot}(0,0,1,1;z) + G^{\mot}(1;z) \zeta_3^{\dR} 
\label{coact.07}
\end{align} \normalsize
A main result of this work is a
streamlined description of the
general form of the
ellipsis in (\ref{coact.06}). In particular, we will employ generating series to formulate  the completion of (\ref{coact.06}) for MPLs in any
% number of variables $a_i \in\{0,1,y,\ldots\}$. 
alphabet, $a_i \in\{0,1,y,\ldots\}$, that excludes the endpoint $z$ of the
integration path.
The complexity of this problem grows rapidly with the 
%number of variables
cardinality of the alphabet. For example,
\small
\begin{align}
&\!\Delta G^{\mot}(1,y;z) = G^{\mot}(1,y;z)  + G^{\mot}(y;z)  G^{\dR}(1;z) + G^{\dR}(1,y;z) 
\notag \\
&\quad
+ G^{\mot}(1;z)  \big[ G^{\dR}(1;y)  - G^{\dR}(0;y) \big]
- G^{\mot}(y;z)  G^{\dR}(1;y)  \label{coact.08} 
\end{align} \normalsize

\medskip
\noindent
{\bf B. $f$-alphabet:}
The so-called {\it $f$-alphabet} \cite{Brown:2011ik, BrownTate} provides a useful description of MZVs 
that exposes the entirety of their currently known relations over $\mathbb Q$. The $f$-alphabet is given by
\begin{itemize}
\item one commutative generator, $f_2$, whose %
$n^{\rm th}$ powers capture
the even Riemann zeta values, $\zeta_{2n} \in \QQ \pi^{2n}$$\!\!$
\vspace{-0.2cm}
\item an infinite set of non-commutative generators, $f_3,f_5,f_7,\ldots$, in odd degrees
$\geq 3$, which capture the odd Riemann zeta values $\zeta_{2k+1}$
\end{itemize}
A word in the $f$-alphabet, $f_2^n f_{i_1}f_{i_2}\ldots f_{i_r}$, for some $n\geq 0$ and any $i_a \in 2 \mathbb{N}+1$, is said to have weight $2n{+}i_1{+}\ldots{+}i_r$. The number of independent words in the $f$-alphabet (recalling that $f_{i_a} f_{i_b} \neq f_{i_b} f_{i_a}$) at fixed weight, $w$, is equal to the expected number of $\mathbb Q$-independent MZVs at weight $w$ \cite{Zagier1994}. 
The space of $f$-polynomials has a multiplication induced by the shuffle product
on the non-commutative generators:
\begin{align}
&(f_2^n f_{i_1} f_{i_2}\ldots f_{i_r}) \shuffle (f_2^{m} f_{i_{r+1}} f_{i_{r+2}} \ldots f_{i_s})
\notag\\
&=  f_2^{m+n}  \sum_{\sigma \in \Sigma(r,s)} f_{i_{\sigma(1)}} f_{i_{\sigma(2)}}\ldots f_{i_{\sigma(r+s)}}
\label{shuffprod} 
\end{align}
Here, the permutations $\sigma \in \Sigma(r,s)$ of $(i_1,\ldots,i_{r+s})$ preserve
the ordering among the first $r$ and the last $s$ elements. E.g. $f_{i_1} \shuffle f_{i_2} = f_{i_1} f_{i_2} {+} f_{i_2} f_{i_1}$.
%The isomorphism $\phi$ into the $f$-alphabet 
%preserves the shuffle product $\phi(\zeta^{\mot}_{\vec{n}}\cdot \zeta^{\mot}_{\vec{u}})
%= \phi(\zeta^{\mot}_{\vec{n}}) \shuffle \phi( \zeta^{\mot}_{\vec{u}})$ (with $\vec{n}, \vec{u} \in \NN^{\times}$), e.g.\ $f_{i} \shuffle f_j = f_i f_j {+} f_j f_i$. The simplest examples read

By formally passing to motivic MZVs\footnote{For the purposes of this work, the
key point of the elaborate definition of motivic MZVs in algebraic-geometry literature
\cite{Goncharov:2005sla, BrownTate, Brown2014MotivicPA, Francislecture} is that they by definition obey no relations other than the currently known $\mathbb Q$-relations among 
conventional MZVs.}, $\zeta_{n_1,\ldots,n_r} \rightarrow \zeta^{\mot}_{n_1,\ldots,n_r}$,
we can isomorphically map MZVs to words in the $f$-alphabet, though this isomorphism is not unique. Any such isomorphism, $\phi$, respects products, in the sense that $\phi(\zeta^{\mot}_{\vec{n}}\cdot \zeta^{\mot}_{\vec{u}})
= \phi(\zeta^{\mot}_{\vec{n}}) \shuffle \phi( \zeta^{\mot}_{\vec{u}})$ (with $\vec{n}, \vec{u} \in \NN^{\times}$), and obeys the following normalization condition at depth one:
\begin{align}
\phi(\zeta^{\mot}_{2k+1}) = f_{2k+1} \, , \ \ \ \ \phi(\zeta^{\mot}_{2}) = f_2
\end{align}
The images of conjecturally indecomposable higher-depth MZVs, such
as $\phi(\zeta^{\mot}_{3,5})$, are determined by imposing that $\phi$ preserves the
motivic coaction. With the notation $^{\mot}$ and $^{\dR}$ to distinguish the
first and second entry of the tensor product, the coaction in the $f$-alphabet
is defined by deconcatenation:
\begin{align}
\Delta (f_2^n f_{i_1}  \ldots f_{i_r})^{\mot} = \sum_{j=0}^r (f_2^n f_{i_1}  \ldots f_{i_j})^{\mot} 
(f_{i_{j+1}}\ldots f_{i_r})^{\dR}
\label{coact.04}
\end{align}
However, this requirement does not
fix the coefficient of $f_{n_1+\ldots+n_r}$ in
$\phi(\zeta^{\mot}_{n_1,\ldots,n_r})$ (with $f_{2n} = \frac{ \zeta_{2n} }{(\zeta_2)^n}f_2^n$
in case of even weight), so the image of $\phi$ beyond depth one
is non-canonical. Following \cite{Schlotterer:2012ny}, our convention 
is to single out the higher-depth MZVs at weight $w \geq 8$ that lead to the conjectural $\mathbb Q$-basis of \cite{Blumlein:2009cf},
\begin{align}
\textrm{basis MZVs} \ni \zeta_{3,5},\, \zeta_{3,7},\, \zeta_{3,3,5}
,\, \zeta_{3,9},\, \zeta_{1,1,4,6}
,\,\ldots
\label{bMZVs}
\end{align}
and to impose $f_{w}$ to be absent in their $\phi$ images. For instance, with this convention, we have
\begin{align}
\phi(\zeta^{\mot}_{3,5}) & = -5 f_3 f_5 \, , \ \ \ \
\phi(\zeta^{\mot}_{3,7})  = - 14 f_3 f_7 - 6 f_5 f_5
\notag\\
\phi(\zeta^{\mot}_{3,3,5}) &= - 5 f_3 f_3 f_5 - 45 f_9 f_2 - \tfrac{6}{5} f_7 f_2^2 + \tfrac{4}{7} f_5 f_2^3
 \label{exfalp}
\end{align}
Note that one can add a $\mathbb Q$-multiple of $f_8$, $f_{10}$ and $f_{11}$ to the expressions for $\phi(\zeta^{\mot}_{3,5})$, $\phi(\zeta^{\mot}_{3,7})$
and $\phi(\zeta^{\mot}_{3,3,5})$ in (\ref{exfalp}) without altering the compatibility of
$\phi$ with products and the coaction. This illustrates that the isomorphism $\phi$ is not canonical and that our choice of the above convention is an ad-hoc choice.

%
% \osnote{We use this conjectural basis for convenience. It is readily available and well-characterized, and moreover its wide use in mathematics and physics allows us to readily cross-check basis-dependent constructions. }

\medskip
\noindent
{\bf C. Single-valued map:} The MPLs of (\ref{coact.01}) are meromorphic and exhibit monodromies when the integration path from 0 to $z$ 
is deformed to wind around the singular points $a_1,\ldots,a_w$ of the integration kernels. 
Still, one can construct single-valued versions ${\rm sv} \, G(a_1,\ldots,a_w;z)$ 
that share the holomorphic differential equations of (\ref{coact.01}) 
\beq
\partial_z  {\rm sv} \, G(a_1,a_2,\ldots,a_w;z)
= \frac{ {\rm sv} \, G(a_2,\ldots,a_w;z) }{z-a_1}
\label{coact.09} 
\eeq
by combining meromorphic MPLs with their complex conjugates and MZVs. 
Explicit constructions of single-valued MPLs in 
%one and multiple variables 
the alphabets $a_i \in \{0,1\}$
and $a_i \in \{0,1,y,\ldots\}$
can be found~in~\cite{svpolylog} and \cite{Broedel:2016kls,DelDuca:2016lad}, respectively.
Generalizations of single-valued MPLs that also accommodate primitives of
$\frac{1}{z\bar z {+} az {+} b\bar z {+} c}$
with $a{,}b{,}c \,{\in} \mathbb C$ are developed in \cite{Schnetz:2021ebf} and go beyond the scope of this work (see \cite{Borinsky:2021gkd, Borinsky:2022lds} for \mbox{applications}).

Similar to the contributions (\ref{coact.06}) to the motivic coaction of MPLs, the generating 
series in \cite{svpolylog, DelDuca:2016lad, Broedel:2016kls} show that the single-valued map includes the following simple class of terms:
\begin{align}
{\rm sv} \, G(a_1,a_2,\ldots,a_w;z)&= \sum_{j=0}^w G(a_1,\ldots,a_j;z) \label{coact.10}  \\
&\quad  \times \overline{ G(a_w,\ldots,a_{j+1} ;z)}
+\ldots \notag
\end{align}
A second main result of this work is a new streamlined formulation of the
additional terms in the ellipsis. The number of terms grows rapidly with the size of the
% number of variables
alphabet, $a_i \in \{0,1,y,\ldots\}$, and
the coefficients of these terms are MZVs or single-valued MPLs
in smaller alphabets.

Similar to (\ref{coact.02}), evaluation at $z=1$ of single-valued MPLs, %in one variable 
with $a_i \in \{0,1\}$,
defines single-valued MZVs 
\cite{Schnetz:2013hqa, Brown:2013gia} \small
\begin{align}
\zeta^{\rm sv}_{n_1,\ldots,n_r}&= (-1)^r {\rm sv} \, G(\underbrace{0,\ldots,0}_{n_r-1},1,\ldots,
\underbrace{0,\ldots,0}_{n_2-1},1,\underbrace{0,\ldots,0}_{n_1-1},1;1)
\label{coact.11} 
\end{align} \normalsize
Single-valued MZVs are expressible in terms of MZVs, but form a proper subset 
since for instance $\zeta^{\rm sv}_{2n}=0$ and
$\zeta^{\rm sv}_{2n+1} = 2 \zeta_{2n+1}$. Examples at higher depth include 
\begin{align}
\zeta^{\rm sv}_{3,5}&= - 10 \zeta_3 \zeta_5 \, , \ \ \ \ 
\zeta^{\rm sv}_{3,7} = - 28 \zeta_3 \zeta_7 - 12 \zeta_5^2 
\label{svmzv.ex} \\
\zeta^{\rm sv}_{3,3,5}&= 2 \zeta_{3,3,5} - 5 \zeta_3^2 \zeta_5
+ 90 \zeta_2 \zeta_9 + \tfrac{12}{5} \zeta_2^2 \zeta_7 - \tfrac{8}{7} \zeta_2^3 \zeta_5
\notag
\end{align} 
The single-valued map, ${\rm sv:} \ \zeta_{n_1,n_2,\ldots,n_r} \rightarrow \zeta^{\rm sv}_{n_1,n_2,\ldots,n_r}$, takes 
the most convenient form
in the $f$-alphabet: 
\beq
{\rm sv} \, f_2^n f_{i_1} f_{i_2}\ldots f_{i_r} = \delta_{n,0} \sum_{j=0}^r f_{i_j}\ldots f_{i_2}f_{i_1} \shuffle f_{i_{j+1}}\ldots f_{i_r} 
\label{coact.12} 
\eeq
This annihilates ${\rm sv}\, f_2 = 0$ and resembles the deconcatenation formula 
(\ref{coact.04}) for the coaction of odd-degree generators $f_{2k+1}$.
Even though the single-valued map is only well-defined in a motivic setting \cite{Brown:2013gia},
we omit the superscripts $\zeta^{\mot}$ and $G^{\mot}$ in a slight abuse of notation, except where they specify the entries of coaction formulas.
Note that the single-valued map preserves the product structure 
${\rm sv}(A\cdot B) = {\rm sv}(A)\cdot {\rm sv}(B)$ for arbitrary 
combination $A,B$ of MPLs and MZVs.

\vspace{\mylength}
%%%%%%%%%%%%%%%%%%%%%%%%%%%%%%%%%%%%%%%%%%%%%%%%%%%%%%%%%
%%%%%%%%%%%%%%%%%%%%%%%%%%%%%%%%%%%%%%%%%%%%%%%%%%%%%%%%%
\section{Key Generating series}
%%%%%%%%%%%%%%%%%%%%%%%%%%%%%%%%%%%%%%%%%%%%%%%%%%%%%%%%%
%%%%%%%%%%%%%%%%%%%%%%%%%%%%%%%%%%%%%%%%%%%%%%%%%%%%%%%%%
\vspace{\myotherlength}

Our results for the motivic coaction and the single-valued map of MPLs 
are best expressed using generating series. In this section, we define the relevant generating series of MPLs and MZVs.

\medskip
\noindent
{\bf A. Polylogarithmic generating series:} Let the alphabet $A$ be a set of
labels, $a \in A$, excluding the endpoint $z$ of the integration path in (\ref{coact.01}), and introduce non-commuting formal variables $e_{a}$ for each label. The meromorphic MPLs are organised into the following generating series
\begin{align}
&\GG_A(z;e_a) = \sum_{r=0}^\infty \! \! \! \!  \sum_{ \ \ a_1,\ldots,a_r \in A} 
\! \! \! \! \! \! \! \! \! \! e_{a_1} \ldots e_{a_r} 
 G(a_r,\ldots, a_2,a_1;z)\! \! \! \! \label{coact.18} \\
&= 1+ \! \sum_{a_1 \in A} \!e_{a_1} G(a_1;z)
+ \! \! \sum_{a_1,a_2 \in A} \! \! e_{a_1} e_{a_2} G(a_2,a_1;z)
+\ldots
\notag 
\end{align}
The generating series $\GG_A(z;e_a)$ is a solution to the following Knizhnik--Zamolodchikov (KZ) equation
\beq
\partial_z \GG_A(z;e_a) = \GG_A(z;e_a) \sum_{a_j\in A} \frac{e_{a_j}}{z-a_j}
\label{specialKZ}
\eeq 
This is a special case of multivariable KZ equation
\beq
\partial_{z_i}\mathbb F(z_a;x_{bc}) =\mathbb F(z_a;x_{bc}) \sum_{j\neq i} \frac{ x_{ij} }{z_i-z_j}
\label{altcoact.18}
\eeq
where the braid operators $x_{ij} = x_{ji}$ obey the
Yang--Baxter relations \cite{jimbo1989introduction,kassel2012quantum} 
\begin{align}
[ x_{ij}, x_{ik}{+}x_{jk} ] = 0 \, , \ \ \ \ 
[ x_{ij}, x_{kl} ] = 0
\label{YBR}
\end{align}
for pairwise distinct $i,j,k,l$. The KZ equation (\ref{specialKZ}) of the polylogarithmic series (\ref{coact.18}) is obtained by setting \mbox{$z_i\mapsto z$}, $z_j\mapsto a_j$, and identifying $x_{ij}$ with $e_{a_j}$ in (\ref{altcoact.18}). Since (\ref{YBR}) implies no relations among the $x_{ij}$ at fixed $i$, the universal enveloping
algebra of the $e_{a_j}$ in (\ref{coact.18}) is freely generated. An individual MPL, $G(a_1,\ldots, a_r;z)$, is uniquely specified by the series $\GG_A(z;e_a)$ 
%as the coefficient of the word $e_{a_r}\cdots e_{a_1}$.
by isolating the word $e_{a_r}\cdots e_{a_1}$.

\medskip
\noindent
{\bf B. Drinfeld associator:} Restricting (\ref{coact.18}) to the alphabet $A = \{0,1\}$ gives the generating series, $\GG_A(z;e_0,e_1)$ of MPLs that depend only on $z$.
% in one variable. 
Evaluating this at $z=1$ yields the Drinfeld associator \cite{Drinfeld:1989st, Drinfeld2}
\begin{align}
\Phi(e_0,e_1) &= 
\sum_{r=0}^\infty  \sum_{a_1,\ldots,a_r \in \{0,1\}} 
\! \! \! \!  e_{a_1} \ldots e_{a_r} G(a_r,\ldots, a_2,a_1;1) \notag\\
&= 1 +\zeta_2 [e_0,e_1] + \zeta_3  \big[[e_0,e_1],e_0{+}e_1\big]  +\ldots
\label{Dassoc}
\end{align}
This is the generating series of shuffle-regularized MZVs, (\ref{coact.02}), with
$G(0;1)=G(1;1)=0$ \cite{LeMura}.
%Denote the coefficients in $\Phi(e_0,e_1)$ of the odd zeta values, $\zeta_{2k+1}$,
%as
%\beq
%W_{2k+1}(e_0,e_1) \equiv \Phi(e_0,e_1) \big|_{ \zeta_{2k+1} }
%\label{defWk}
%\eeq 
%Because of the multitude of $\QQ$ relations among MZVs, coefficients of MZVs in $\Phi(e_0,e_1)$ are non-canonical. Our convention is that all the MZVs at weight $2k{+}1$ in $\Phi(e_0,e_1)$ are to be decomposed into the conjectural $\mathbb Q$-basis with higher-depth elements in (\ref{bMZVs}) and \cite{Blumlein:2009cf}. With this understood, $W_{2k+1}$ is a unique Lie polynomial in~$e_i$.
Suppose the MZVs in the series $\Phi(e_0,e_1)$ are decomposed into the conjectural $\mathbb Q$-basis with higher-depth elements in (\ref{bMZVs}) \cite{Blumlein:2009cf}. Then, each basis MZV $\zeta_{\vec{n}}$ at weight $w$ is accompanied by a rational degree-$w$ polynomial in $e_0$, $e_1$, which we will refer to as {\it coefficients} of the basis MZVs to be denoted by $\Phi(e_0,e_1)|_{\zeta_{\vec{n}}}$. At odd weights, $w = 2k{+}1$, the coefficients of $\zeta_{2k+1}$ obtained with our choice of basis are Lie polynomials $W_{2k+1}(e_0,e_1)$ of degree $2k{+}1$
\beq
W_{2k+1}(e_0,e_1) \equiv
\Phi(e_0,e_1) |_{\zeta_{2k+1}}
\label{defWk}
\eeq 
Starting from weight $2k{+}1=11$, these $W_{2k+1}(e_0,e_1)$ are non-canonical, but specified by our ad-hoc choice to fix the MZV basis (\ref{bMZVs}). This freedom to pick a basis is the same as the freedom to choose the isomorphism $\phi$.

\medskip
\noindent
{\bf C. Zeta generators:} For the results in this paper, it is important to introduce a second generating
series of MZVs, $\MM$, adapted to the $f$-alphabet. Introducing
non-commuting formal generators $M_{3},M_5,\ldots$, we define
\begin{align}
\MM &= \sum_{r=0}^\infty    \sum_{ i_1,\ldots,i_r \in 2\NN+1}  \! \! \! 
 \phi^{-1}(f_{i_1}  \ldots f_{i_r}) M_{i_1}   \ldots M_{i_r}
\label{coact.15}  \\
&= 1+  \! \! \!  \sum_{i_1 \in 2\NN+1}  \! \! \! \zeta_{i_1} M_{i_1} 
+ \! \! \!  \! \! \!  \sum_{i_1,i_2 \in 2\NN+1}  \! \! \!  \! \! \!  \phi^{-1}(f_{i_1} f_{i_2} )M_{i_1} M_{i_2}
+\ldots
\notag
\end{align}
The \emph{zeta generators}, $M_{2k+1}$ \cite{DG:2005, Brown:depth3},
have no relations among themselves and cannot be expressed in terms of braid 
operators $e_{a_i}$. However, we will later encounter an algebra on the $M_{2k+1}$ and $e_{a_i}$, such that commutators $[M_{2k+1},e_{a_i}]$ are equal to 
nested commutators of the braid operators, $e_{a_i}$. In other words, the
braid algebra is normalized by the $M_{2k+1}$.

However, recall that the $f$-alphabet of MZVs is non-canonical. For instance,
departing from our convention specified in the discussion around (\ref{exfalp})
would redefine
$M_{11}$ by adding a rational multiple of $[[M_3,M_5],M_3]$.
Similar ambiguities apply to $M_{13},M_{15},\ldots$, which can
be redefined by nested brackets of an odd number of $M_{2k+1}$.

The generating series $\MM$ is related to similar series encountered in open-string tree-level amplitudes
\cite{Schlotterer:2012ny} and configuration-space integrals
over marked points on the sphere \cite{Britto:2021prf}. However, in these contexts, 
one studies specific matrix representations of the Lie algebra satisfied by the $M_{2k+1}$ and $e_{a_i}$. It is important for our results
to refrain from specifying a matrix representation. In this way, we ensure that no terms in (\ref{coact.15}) drop out due to non-generic relations that can arise in specific matrix representations.

\medskip
\noindent
{\bf D. Inverse series:} 
Both polylogarithms and the $f$-alphabet algebra obey the 
shuffle relations. This makes it straightforward to verify
that the series inverses of $\GG_A$ and $\MM$ with respect to concatenation are
\begin{align}
&\GG_A(z;e_a)^{-1} = \sum_{r=0}^\infty (-1)^r \! \! \! \!    \sum_{a_1,\ldots,a_r \in A} 
\! \! \! \! e_{a_1} \ldots e_{a_r} 
 G(a_1,\ldots, a_r;z)   \notag\\
 &\MM^{-1} = \sum_{r=0}^\infty (-1)^r \! \! \! \! \! \! \! \! \!  \!\sum_{i_1,i_2,\ldots,i_r \in 2\NN+1}\! \! \!  \! \! \! \! \! \! \! 
\phi^{-1}( f_{i_1} f_{i_2} \ldots f_{i_r} )M_{i_r}  \ldots M_{i_2}M_{i_1}
 \label{invser}
 \end{align}

\vspace{\mylength}
%%%%%%%%%%%%%%%%%%%%%%%%%%%%%%%%%%%%%%%%%%%%%%%%%%%%%%%%%
%%%%%%%%%%%%%%%%%%%%%%%%%%%%%%%%%%%%%%%%%%%%%%%%%%%%%%%%%
\section{Polylogarithms in one variable}

%%%%%%%%%%%%%%%%%%%%%%%%%%%%%%%%%%%%%%%%%%%%%%%%%%%%%%%%%
%%%%%%%%%%%%%%%%%%%%%%%%%%%%%%%%%%%%%%%%%%%%%%%%%%%%%%%%%
\vspace{\myotherlength}

We begin by presenting our results as applied to MPLs with the alphabet
$a_i \in \{ 0,1\}$, to be referred to as {\it MPLs in one variable}. The generalisation to larger alphabets
(or `MPLs in more than one variable') is given
in later sections. Recall that the generating series for MPLs in one
variable is $\GG_{\{0,1\}}(z) =\GG_{\{0,1\}}(z;e_0,e_1) $,
where the alphabet in (\ref{coact.18}) is specialized to $A = \{0,1\}$.

\medskip
\noindent
{\bf A. The motivic coaction:} Our first main result is a formula for
the motivic coaction. We claim that the motivic coaction on the generating series is given by
\beq
\Delta \GG^{\mot}_{\{0,1\}}(z) = (\MM^{\dR})^{-1} \,\GG^{\mot}_{\{0,1\}}(z)  \, \MM^{\dR}\,
\GG^{\dR}_{\{0,1\}}(z) 
\label{coact.36}
\eeq
where the coaction $\Delta$ of individual MPLs in one variable is uniquely specified by the coefficient
of $e_{a_r}\ldots e_{a_2}e_{a_1}$ in the generating series $\Delta \GG^{\mot}_{\{0,1\}}(z)$,\footnote{While the restriction $|_{\zeta_{2k+1}}$ in (\ref{defWk}) isolates words in $e_0, e_1$ multiplying
$\zeta_{2k+1}$ after basis decompositions specified above, the restriction $|_{e_{a_r}\ldots e_{a_2}e_{a_1}}$ to words in free-Lie-algebra generators $e_a$ of (\ref{pickupmot}), (\ref{pickupsv}) and similar formulae below isolates combinations of MPLs and MZVs. In both cases, the results of the restrictions $|_{\zeta_{2k+1}}$ and $|_{e_{a_r}\ldots e_{a_2}e_{a_1}}$ are referred to as \it{coefficients}.}
\beq
\Delta G^{\mot}(a_1,\ldots ,a_r;z) =
\Delta \GG^{\mot}_{\{0,1\}}(z) |_{e_{a_r}\ldots e_{a_2}e_{a_1}}
\label{pickupmot}
\eeq
The `motivic' and `de Rham' superscripts of $\GG^{\mot},\GG^{\dR},\MM^{\dR}$ apply to the MPLs and MZVs in
the respective expansions (\ref{coact.18}), (\ref{coact.15}),
and they distinguish the first and second entry of the series $\Delta \GG^{\mot}_{\{0,1\}}(z)$.
The multiplication of the generating series in (\ref{coact.36}) is just
the concatenation product of the braid operators $e_0,e_1$
and the zeta generators $M_{2k+1}$.

As a first check of (\ref{coact.36}), note that the simple terms (\ref{coact.06}) in the coaction formula for MPLs
are recovered by ignoring all the MZV terms in $\MM$, i.e.\ setting $\MM^{\dR} \rightarrow 1$ in (\ref{coact.36}). The additional terms involving MZVs arise in our formula (\ref{coact.36}) from commutators of the $M_{2k+1}$ and $e_{a}$. This algebra
is specified below in subsection {\bf C}.

\medskip
\noindent
{\bf B. The sv map:} Our second main result is a formula for the single-valued map, ${\rm sv}$. We claim that the ${\rm sv}$ map on MPLs in one variable is given by
\beq
\!{\rm sv}\, \GG_{\{0,1\}}(z) = ({\rm sv}\,\MM)^{-1} \,\overline{ \GG_{\{0,1\}}(z)^t }
\, ({\rm sv}\, \MM) \,  \GG_{\{0,1\}}(z)
\label{coact.57}
\eeq
where the action of ${\rm sv}$ on individual MPLs is uniquely specified by the coefficient
\beq
{\rm sv}\,G(a_1,a_2,\ldots ,a_r;z)
={\rm sv}\, \GG_{\{0,1\}}(z)  \big|_{e_{a_r}\ldots e_{a_2}e_{a_1}}
\label{pickupsv}
\eeq
of the generating series ${\rm sv}\, \GG_{\{0,1\}}(z)$.

The factor $\overline{ \GG_{\{0,1\}}(z)^t }$ in (\ref{coact.57}) is defined
from the series $\GG_{\{0,1\}}(z)$ by applying complex conjugation to the MPLs and 
word reversal on the braid operators:
$(e_{a_1} \cdots e_{a_r})^t = e_{a_r} \cdots e_{a_1}$. 
The factor ${\rm sv}\,\MM$ in (\ref{coact.57}) is a generating series of single-valued
MZVs. This is defined from the series expansion (\ref{coact.15}) of $\MM$ 
by acting with the ${\rm sv}$ on the $f$-algebra, (\ref{coact.12}).
For example, the first few terms are
\begin{align}
{\rm sv}\,\MM &= 1+ 2 \! \! \!  \sum_{i_1 \in 2\NN+1}  \! \! \! \zeta_{i_1} M_{i_1} 
+ 2 \! \! \!  \! \! \!  \sum_{i_1,i_2 \in 2\NN+1}  \! \! \!  \! \! \!  \zeta_{i_1} \zeta_{i_2} M_{i_1} M_{i_2}  \label{simpsvm} \\
&\quad + \! \! \!  \! \! \! \! \! \!    \sum_{i_1,i_2,i_3 \in 2\NN+1} \! \! \!  \! \! \!  \! \! \!  
\phi^{-1} \big( {\rm sv}(f_{i_1} f_{i_2} f_{i_3}) \big) M_{i_1}M_{i_2} M_{i_3}
+\ldots
\notag
\end{align}
The factors of 2 arise from ${\rm sv}(f_i) = 2f_i$ and ${\rm sv}(f_if_j) = 2f_i \shuffle f_j$, and $\phi^{-1} ( {\rm sv}(f_{i_1} f_{i_2} f_{i_3}) )$ at weight $i_1{+}i_2{+}i_3 \geq 11$ comprise (conjecturally) irreducible single-valued MZVs at depth $\geq 3$, see for instance 
(\ref{exfalp}) and (\ref{svmzv.ex}).

As a first check of (\ref{coact.57}), note that the simple terms (\ref{coact.10}) in the expression for single-valued MPLs are recovered 
by setting ${\rm sv}\,\MM \rightarrow 1$ in (\ref{coact.57}), i.e.\ ignoring all MZVs.

\medskip
\noindent
{\bf C. Commutators with zeta generators:} Both of the new formulas, (\ref{coact.36}) and (\ref{coact.57}), for the motivic coaction and ${\rm sv}$ map of MPLs in one variable involve conjugations by a series in MZVs and zeta generators $M_{2k+1}$. In order to attain well-defined coefficients of $e_{a_r} \ldots e_{a_1}$ in (\ref{pickupmot}) and (\ref{pickupsv}), we must specify how conjugations with $M_{2k+1}$ conspire with the braid operators to yield a series solely in~$e_i$.

As a first step, we expand the conjugation of an
arbitrary series $\GG$ in the braid operators $e_0,e_1$
with $\MM$ in terms of nested commutators
\begin{align}
&\phi \big( (\MM )^{-1} \mathbb G  \MM  \big)= 
\sum_{r=0}^\infty  \sum_{i_1,i_2,\ldots,i_r \in 2\NN+1}  \! 
f_{i_1} f_{i_2} \ldots f_{i_r}   \label{coact.37}  \\
 &\quad \quad\quad\quad\quad\quad\quad \times \big[ \big[\ldots [[\mathbb G, M_{i_1}], M_{i_2}],\ldots
,M_{i_{r-1}} \big],M_{i_r}\big]
\notag  \\
&= \mathbb G+  \! \! \! \! \sum_{i_1 \in 2\NN+1}  \!  \! \! \! f_{i_1} [\mathbb G,M_{i_1} ] 
+ \! \! \! \! \! \!\! \sum_{i_1,i_2 \in 2\NN+1} \!\! \! \! \! \! \!  f_{i_1} f_{i_2}  \big[[\mathbb G,M_{i_1}], M_{i_2} \big]
+\ldots
\notag
\end{align}
The second step is to specify the following commutation relations among the zeta generators $M_{2k+1}$ that 
normalize the free algebra generated by $e_0,e_1$
\beq
[e_0 , M_{2k+1} ] = 0 \, , \ \ \ \
%[e_1 , M_{2k+1}] = \big[ \Phi(e_0,e_1) \big|_{\zeta_{2k+1}} , e_1 \big]
[e_1 , M_{2k+1}] = \big[W_{2k+1}(e_0,e_1)  , e_1 \big]
\label{coact.40} 
\eeq
where $W_{2k+1}(e_0,e_1)$ are the Lie polynomials appearing in the expansion of the
Drinfeld associator, (\ref{defWk}). These relations will be motivated in subsection \textbf{D}, below. They are uniquely determined by $\mathbb Q$-relations among MZVs at weight $2k{+}1$, together with the basis choice that allows us to fix the $W_{2k+1}(e_0,e_1)$. For example,
identifying $W_{3}(e_0,e_1) = [[e_0,e_1],e_0{+}e_1] $ from~(\ref{Dassoc}),
\begin{align}
[e_0 , M_{3}] = 0 \, , \ \ \ \
[e_1 , M_{3}] = \big[ \big[ [e_0,e_1],e_0{+}e_1\big] ,e_1\big]   \label{coact.41a} 
\end{align} 
These relations featured in the context of the stable derivation algebra\footnote{For instance, $W_3$ satisfies the equation for commutators \emph{`f'} in the formulation of standard derivation algebra as stated in section 2.2 of \cite{Furusho2000TheMZ}. Taking the commutators in (\ref{coact.41a}) with respect to $M_3$ gives rise to the corresponding derivation $D_{W_3}$ on the free Lie algebra generated by $e_0,e_1$.} \cite{Ihara:stable, Furusho2000TheMZ} and were later observed for specific matrix representations 
in (4.23) of \cite{Britto:2021prf}.

Upon iterative use of the commutators (\ref{coact.40}) in (\ref{coact.37}), all the zeta generators entering (\ref{coact.36}) and (\ref{coact.57}) via $(\MM^{\dR})^{-1} \GG^{\mot}_{\{0,1\}}(z)  \MM^{\dR}$ and $({\rm sv}\,\MM)^{-1} \overline{ \GG_{\{0,1\}}(z)^t }({\rm sv}\, \MM)$ conspire to yield series $\Delta \GG^{\mot}_{\{0,1\}}(z)$ and ${\rm sv}\, \GG_{\{0,1\}}(z)$ solely in  $e_0$ and $e_1$.
This is a crucial prerequisite to extract the motivic coaction and single-valued map of MPLs in one variable from the respective coefficients in (\ref{pickupmot}) and (\ref{pickupsv}).

For example, the simplest contribution to the motivic coaction (\ref{coact.36}) involving a MZV from $\MM$ is
\begin{align}
&(\MM^{\dR})^{-1} \GG_{\{0,1\}}(z)  \MM^{\dR} \big|_{\zeta^{\dR}_3}  \label{exzeta3} 
 \\
&\quad = 
G(1;z) \big[ \big[ [e_0,e_1],e_0{+}e_1\big] ,e_1\big] + \ldots
\notag
\end{align}
with MPLs of weight $\geq 2$ and words involving $\geq 5$ braid operators in the ellipsis. Isolating the word $e_1e_1e_0e_0$ in (\ref{exzeta3}), this reproduces the
contribution $G^{\mot}(1;z)\zeta^{\dR}_3$ to $\Delta G^{\mot}(0,0,1,1;z)$ in
the introductory example (\ref{coact.07}).

Likewise,
the simplest MZV contribution to (\ref{coact.57}) is
\begin{align}
&({\rm sv}\,\MM)^{-1} \,\overline{ \GG_{\{0,1\}}(z)^t }
\, ({\rm sv}\, \MM) \big|_{\zeta_3} \label{svexzeta3} \\
&\ \  = 2 \overline{ G(1;z)}  \big[ \big[ [e_0,e_1],e_0{+}e_1\big] ,e_1\big] + \ldots
\notag
\end{align}
where the ellipsis contains words with $\geq 5$ braid operators. Isolating the
word $e_1e_1e_0e_0$, this reproduces the contribution $2 \zeta_3  \overline{ G(1;z)}$ to ${\rm sv} \, G(0,0,1,1;z)$.

\medskip
\noindent
{\bf D. Equivalence with earlier results:} We claim that the formula 
(\ref{coact.36}) for the motivic coaction on MPLs in one variable is 
equivalent to the Ihara formula \cite{Ihara1989TheGR}
\begin{align}
\Delta \GG^{\mot}_{\{0,1\}}(z;e_0,e_1) &= \GG^{\mot}_{\{0,1\}} (z;e_0,e_1' ) \GG^{\dR}_{\{0,1\}}(z;e_0,e_1) \notag\\
e_1' &= \Phi^{\dR}(e_0,e_1) e_1 \Phi^{\dR}(e_0,e_1)^{-1}
\label{oldmot}
\end{align}
which is implicit in (6.6) of \cite{Brown:2013gia} and explicit in Proposition 8.3 of \cite{Brown:2019jng}. Similarly, we claim that the formula (\ref{coact.57}) for the generating series of single-valued MPLs is equivalent to the following
construction of Brown \cite{svpolylog}
\begin{align}
{\rm sv}\,  \GG_{\{0,1\}}(z;e_0,e_1) &= \overline{  \GG_{\{0,1\}}(z;e_0,\hat e_1)^t }\,  \GG_{\{0,1\}}(z;e_0,  e_1) \notag \\
\hat e_1 &= \big( {\rm sv}\,  \Phi(e_0,e_1)\big)\, e_1 \,\big( {\rm sv}\, \Phi(e_0,e_1)\big)^{-1} \notag \\
&= {\rm sv} \, e_1'
\label{oldsv}
\end{align}
Unlike our expressions, (\ref{coact.36}) and (\ref{coact.57}), the earlier formulas, (\ref{oldmot}) and (\ref{oldsv}), employ a
change of alphabet $e_1 \rightarrow e_1'$ or $\hat e_1$ for one of the braid
operators within the series $\GG^{\mot}_{\{0,1\}}(z)$ or $ \overline{  \GG_{\{0,1\}}(z)^t }$.

A full proof that our results follow from these earlier formulas is to be given in
\cite{Frost:2024}. Here we sketch the argument in brief. First, insert
$1= \MM^{\dR} (\MM^{\dR})^{-1}$ and $1= {\rm sv}\, \MM ({\rm sv}\,\MM)^{-1}$ between any
pair of braid operators $e_i$ in the series $\GG_{\{0,1\}}^{\mot}(z)$ in (\ref{coact.36}) 
and $\overline{ \GG_{\{0,1\}}(z)^t }$ in (\ref{coact.57}), respectively. 
Given that $e_0$ commutes with the $M_{2k+1}$,
\begin{align}
(\MM^{\dR})^{-1} e_0 \,\MM^{\dR} 
= ({\rm sv}\,\MM)^{-1}e_0 \,({\rm sv}\, \MM) = e_0
\end{align}
our formulas, (\ref{coact.36}) and (\ref{coact.57}), can be rewritten as
\begin{align}
\Delta \GG^{\mot}_{\{0,1\}}(z;e_0,e_1) &=  \GG^{\mot}_{\{0,1\}}\big(z;e_0,(\MM^{\dR})^{-1} e_1 \MM^{\dR} \big)  \notag \\
&\quad \quad \times
\GG^{\dR}_{\{0,1\}}(z;e_0,e_1)  \label{altreform} \\
%%%
{\rm sv}\, \GG_{\{0,1\}}(z;e_0,e_1) &= \overline{ \GG_{\{0,1\}}\big(z;e_0, ({\rm sv}\,\MM)^{-1}e_1 {\rm sv}\, \MM \big)^t } \notag \\
&\quad \quad \times
\,   \GG_{\{0,1\}}(z;e_0,e_1) \notag
\end{align} 
Next, it remains to show that
\beq
(\MM^{\dR})^{-1} e_1 \MM^{\dR} =  \Phi^{\dR}(e_0,e_1) e_1 \Phi^{\dR}(e_0,e_1)^{-1}
\label{lefttoshow}
\eeq
As a first check, note that the coefficients of $\zeta^{\dR}_{2k+1}$ on both
sides of this equation agree, as a consequence of the commutators (\ref{coact.40}). 
In the full proof of \eqref{lefttoshow}, to be given in \cite{Frost:2024}, matching the appearance of $f_{i_1}\ldots f_{i_r}$ at $r\geq 2$ relies on the coaction properties of $\Phi(e_0,e_1)$ and~$\MM$.

Note that the changes of alphabet in the earlier formulas, (\ref{oldmot}) and (\ref{oldsv}), cannot be rewritten by conjugating
$\GG^{\mot}_{\{0,1\}}(z)$ or $ \overline{  \GG_{\{0,1\}}(z)^t }$ with the Drinfeld
associator, $\Phi^{\dR}(e_0,e_1)$ or ${\rm sv}\, \Phi(e_0,e_1)$. This is because the Drinfeld associator does not commute with $e_0$. Our formulas are only made possible by introducing the zeta generators $M_{2k+1}$ outside the universal enveloping algebra of $e_0, e_1$.

\medskip
\noindent
{\bf E. Correlations between different MZVs:} Our formulas, (\ref{coact.36}) and (\ref{coact.57}), for the motivic coaction and the single-valued
map, expose a key fact. In both cases, the 
%coefficients 
appearance in these generating series of higher-depth MZVs $\phi^{-1}(f_{i_1} \ldots f_{i_r} )$ ($r\geq 2$) is completely determined by the appearance of the odd zeta values, $\zeta_i$.
%are completely determined by the coefficients of the odd zeta values, $\zeta_i=\phi^{-1}(f_i)$. 
%
This correlation between MZVs of different depth is closely related to the
simple properties
\beq
\Delta \MM^{\mot} = \MM^{\mot} \MM^{\dR}
\, , \ \ \ \
{\rm sv} \, \MM  = \MM^t \MM
\label{coact.17}
\eeq
of the series $\MM$ and can be understood from the formal similarity of (\ref{coact.15}) with a path-ordered exponential of the sum over combinations $f_{2k+1}M_{2k+1}$. In the adjoint actions of $\MM$ in (\ref{coact.36}) and (\ref{coact.57}) the zeta generators act via the basis-dependent Lie polynomials $W_{2k+1}$ in (\ref{coact.40}) and generate the motivic Lie algebra. 
% \osnote{and follows from the fact that picking a $\mathbb Q$ basis of MZVs specifies Lie polynomials $W_{2k+1}$ that determine the action of zeta generators $M_{2k+1}$}. 
By contrast, the 
Drinfeld associator does not satisfy 
%the properties (\ref{}).
any direct analogues of (\ref{coact.17}).
The expressions for $\Delta \Phi^{\mot}(e_0,e_1)$ and ${\rm sv} \, \Phi(e_0,e_1)$
\cite{Drummond:2013vz, Brown:2013gia} necessitate the above change of 
alphabet $e_1 \rightarrow e_1'$ or $\hat e_1$.

The proof of (\ref{lefttoshow}) in \cite{Frost:2024} follows from understanding
how these properties of $\MM$ and $\Phi(e_0,e_1)$ are related. The 
Drinfeld associator enjoys an expansion similar to that of $\MM$ in (\ref{coact.15}), in which the
concatenation product on the zeta generators $M_{2k+1}$ is replaced by a certain product $\circ$
on the Lie polynomials $W_{2k+1}$ \cite{Deligne1989TheGR,Drummond:2013vz}.

\vspace{\mylength}
%%%%%%%%%%%%%%%%%%%%%%%%%%%%%%%%%%%%%%%%%%%%%%%%%%%%%%%%%
%%%%%%%%%%%%%%%%%%%%%%%%%%%%%%%%%%%%%%%%%%%%%%%%%%%%%%%%%
\section{Polylogarithms in two variables}
%%%%%%%%%%%%%%%%%%%%%%%%%%%%%%%%%%%%%%%%%%%%%%%%%%%%%%%%%
%%%%%%%%%%%%%%%%%%%%%%%%%%%%%%%%%%%%%%%%%%%%%%%%%%%%%%%%%
\vspace{\myotherlength}

The results above for MPLs in one variable (i.e.\ MPLs with the alphabet $a_i \in \{0,1\}$) generalize to larger alphabets. 
%$A= \{ 0,1,y_1,\ldots,y_r\}$ 
%excluding $z$, where the associated 
We consider MPLs $G(a_1,\ldots,a_w;z)$ with alphabets $a_i \in \{ 0,1,y_1,\ldots,y_r\}$ that exclude $z$ (i.e.\ $y_i\neq z$) and refer to these as MPLs in $(r{+}1)$ variables.
Before presenting the general result in the next section, we devote this
section to the case of MPLs in two variables, i.e.\ $a_i \in \{ 0,1,y\}$.
In this way, the additional features of the Lie algebras governing the multivariable case (as opposed to MPLs in one variable) will be illustrated in a combinatorially simple setting.

We study MPLs in two variables, $y$ and $z$, in the fibration basis involving products of MPLs $G(a_1,\ldots,a_r;z)$ ($a_i \in \{0,1,y\}$) and $G(b_1,\ldots,b_s;y)$ ($b_i \in \{0,1\}$). The respective generating series 
for these MPLs are now defined using two 
sets of braid-group generators. To avoid confusion, we use the superscript $^{(2)}$ to distinguish the new generators from the braid operators $e_0,e_1$ of the one-variable case. Hence,
the protagonist of this section is the generating series (\ref{coact.18}) with $A=\{0,1,y\}$,
\beq
\GG_{\{0,1,y\}}(z) = \GG_{\{0,1,y\}}(z;e_0^{(2)},e_1^{(2)},e_y^{(2)})
\label{2varGG}
\eeq

\medskip
\noindent
{\bf A. The results:} For MPLs in two variables, we claim that the motivic
coaction is generated by the formula
\begin{align}
\Delta \GG^{\mot}_{\{0,1,y\}}(z) &= \big(\GG^{\dR}_{\{0,1\}}(y) \big)^{-1}
\, ( \MM^{\dR}  )^{-1} \, \GG^{\mot}_{\{0,1,y\}}(z) \notag \\
&\quad \times \MM^{\dR}\, \GG^{\dR}_{\{0,1\}}(y)  \,
\GG^{\dR}_{\{0,1,y\}}(z) \label{coact.42}
\end{align}
Similarly, for MPLs in two variables, our formula for the single-valued map is
\begin{align}
{\rm sv}\, \GG_{\{0,1,y\}}(z) &= \big({\rm sv}\, \GG_{\{0,1\}}(y) \big)^{-1}
\, ({\rm sv}\, \MM  )^{-1} \, \overline{ \GG_{\{0,1,y\}}(z)^t} \notag \\
&\quad \times ({\rm sv}\, \MM)\, \big({\rm sv}\,\GG_{\{0,1\}}(y)\big)  \,
\GG_{\{0,1,y\}}(z) \label{svcoact.42}
\end{align}
see (\ref{simpsvm}) and (\ref{coact.57})
for the generating series ${\rm sv}\,\MM$ and ${\rm sv}\,\GG_{\{0,1\}}(y)$ of single-valued MZVs and one-variable MPLs, respectively.

The action of $\Delta$ and ${\rm sv}$ on a specific MPL $G(a_1,\ldots,a_r;z)$ in two variables ($a_i \in \{0,1,y\}$) is uniquely specified by extracting the coefficient of the word $e_{a_r}^{(2)}\cdots e_{a_1}^{(2)}$ in (\ref{coact.42}) and (\ref{svcoact.42}), respectively.
However, both (\ref{coact.42}) and (\ref{svcoact.42}) feature zeta generators $M_{2k+1}$ and the distinct braid operators $e_0,e_1$ of the one-variable case on the right-hand side.
To use these formulas, one must eliminate all of $M_{2k+1},e_0,e_1$ in favor of
the generators $e_0^{(2)},e_1^{(2)},e_y^{(2)}$ on the left-hand side. 
This can be done by using
the Yang--Baxter relations of the braid algebra, (\ref{YBR}), and also extending
the braid algebra by specifying the commutators of the $M_{2k+1}$ with the braid 
operators.

\medskip
\noindent
{\bf B. The braid algebra:} In the present notation, the braid-algebra relations for two variables are given by
\begin{align}
[e^{(2)}_0, e_0] &= 0\, ,
&[  e^{(2)}_0, e_1] &= 0 \label{coact.44} \\
[  e^{(2)}_1, e_0] &= [e^{(2)}_1, e^{(2)}_{y}] \, ,
&[  e^{(2)}_1, e_1] &= [e^{(2)}_{y},e^{(2)}_1] \notag \\
[  e^{(2)}_{y}, e_0] &=  [e^{(2)}_0, e^{(2)}_{y}]  \, ,
&[  e^{(2)}_{y}, e_1] &= [e^{(2)}_1, e^{(2)}_{y}]
\notag
\end{align}
These commutators can be derived from the Yang--Baxter relations (\ref{YBR}) under the identifications
\begin{align}
e_0^{(2)} &= x_{z0} \, , \ \ \ \ e_1^{(2)} = x_{z1}  \, , \ \ \ \ e_y^{(2)} = x_{yz}
\notag \\
e_0 &= x_{y0}{+}x_{yz}  \, , \ \ \ \  e_1 = x_{y1}
\label{identYB}
\end{align}
which follow from the KZ equation (\ref{altcoact.18}) of
the product, $\GG_{\{0,1\}}(y)   \GG_{\{0,1,y\}}(z)$, as in \cite{Britto:2021prf}.

The key point is that the braid-algebra relations (\ref{coact.44}) do not feature any operators $e_0,e_1$ of the one-variable case on their right-hand side. One can therefore eliminate any appearance of $e_0,e_1$ in favor of $e_i^{(2)}$ on the right-hand sides of our main formulas (\ref{coact.42}) and (\ref{svcoact.42}). More specifically, iterative use of (\ref{coact.44}) ensures that the expansion~of
\begin{align}
&\GG_{\{0,1\}}(y)^{-1} \, \mathbb X \, \GG_{\{0,1\}}(y) = \mathbb X 
+ \sum_{a_1 \in \{0,1\}} G(a_1;y) [  \mathbb X , e_{a_1} ] \notag \\
&\quad\quad\quad  +\sum_{a_1,a_2 \in \{0,1\}} G(a_1,a_2;y) \big[ [  \mathbb X , e_{a_2} ],e_{a_1} \big] 
+ \ldots
\label{conviaGG}
\end{align}
preserves the alphabet $\{e_0^{(2)},e_1^{(2)},e_y^{(2)}\}$ for any series $\mathbb X$ in these three variables.

\medskip
\noindent
{\bf C. The extension by zeta generators:}
Finally, to complete the presentation of the main results (\ref{coact.42}) and (\ref{svcoact.42}) in terms of $e_0^{(2)},e_1^{(2)},e_y^{(2)}$, we extend the braid algebra (\ref{coact.44}) by the following commutators for zeta generators:
\begin{align}
[e_{0}^{(2)}, M_{2k+1}]  &= 0 \label{coact.49} \\
[e_{y}^{(2)}, M_{2k+1}]  &= \big[ W_{2k+1}(e_0^{(2)},e_{y}^{(2)})  , e_{y}^{(2)} \big]
\notag \\
[e_1^{(2)}, M_{2k+1}]  &=
 \big[ W_{2k+1}(e_0^{(2)}{+}e_{y}^{(2)},e_1^{(2)})
{+}W_{2k+1}(e^{(2)}_0,e^{(2)}_{y})\notag \\
 &\quad \quad
{+} W_{2k+1}(e_{y}^{(2)},e_0{-}e^{(2)}_{y}), e^{(2)}_1 \big]
\notag
\end{align} 
\normalsize
Again, $W_{2k+1}$ are the Lie-polynomial coefficients (\ref{defWk}) of the
series expansion (\ref{Dassoc}) of the Drinfeld associator. These commutators involving $e^{(2)}_{i}$ generalize the Lie brackets (\ref{coact.40}) with zeta generators in the one-variable case. 
The derivation of (\ref{coact.49}) will be given in \cite{Frost:2024} and is 
based on analytic continuations of the series $\GG_{\{0,1,y\}}(z)$ in two-variable MPLs.
Note that the braid-algebra relations (\ref{coact.44}) can be used
to eliminate the $e_0$ from the Lie polynomial $W_{2k+1}(e_{y}^{(2)},e_0{-}e^{(2)}_{y})$ in the last line of (\ref{coact.49}).

The relations (\ref{coact.49}) appear new to the literature. They are not immediate consequences of the stable derivation algebra of Ihara but still have indirect connections, as will be explained in the proof of these relations in \cite{Frost:2024}.

The extended braid algebra in (\ref{coact.49}) and (\ref{coact.44}) can now be used to extract the coaction and ${\rm sv}$ map of individual MPLs in two variables from their generating series (\ref{coact.42}) and (\ref{svcoact.42}). The first step is to conjugate by $\MM^{\dR}$ (or by ${\rm sv}\,\MM$) via (\ref{coact.37}), and to then eliminate
the $M_{2k+1}$ from the series using the commutators (\ref{coact.49}). The second
step is to conjugate by $\GG^{\dR}_{\{0,1\}}(y)$ (or by ${\rm sv}\,\GG_{\{0,1\}}(y)$)  via (\ref{conviaGG}), and to then eliminate the $e_0,e_1$ using the braid relations~(\ref{coact.44}).

The introductory example (\ref{coact.08}) for $\Delta G^{\mot}(1,y;z)$ can be extracted from the coefficient of $e_y^{(2)}e_1^{(2)}$ in the coaction formula (\ref{coact.42}). The weight-one 
contributions from $\GG^{\mot}_{\{0,1,y\}}(z)$ on the 
right-hand side suffice to recover the second line of (\ref{coact.08}) from
the conjugation (\ref{conviaGG}) and the braid algebra (\ref{coact.44}).

\medskip
\noindent
{\bf D. Equivalence with earlier results:}
The proof of our results, (\ref{coact.42}) and (\ref{svcoact.42}), again follows
by showing that they are equivalent to earlier formulas. In \cite{Frost:2024}, it
will be shown that our coaction formula (\ref{coact.42}) is equivalent to the multivariate generalization of the Ihara
formula in \cite{Brown:2019jng}, and that our formula for the ${\rm sv}$ map (\ref{svcoact.42}) is
equivalent to the results in \cite{DelDuca:2016lad} for single-valued MPLs in
two variables.

As in the single-variable case, these earlier formulas introduce a change of 
alphabet, defining new letters such as
$\hat e_1^{(2)},\hat e_y^{(2)}$ in case of the single-valued map \cite{DelDuca:2016lad}. The first step of the proof
is to show that the conjugation by $ \MM^{\dR}\, \GG^{\dR}_{\{0,1\}}(y)$ in (\ref{coact.42}) 
and by ${\rm sv}(\MM \GG_{\{0,1\}}(y))$ in (\ref{svcoact.42}) can be rewritten
as a change of
alphabet from $e_0^{(2)}, e_1^{(2)}, e_y^{(2)}$ to the new alphabet of \cite{Brown:2019jng} and \cite{DelDuca:2016lad}. Note that $e_0^{(2)}$ is not changed
in this substitution, and this is consistent with the fact that $e_0^{(2)}$
commutes with $M_{2k+1},e_0,e_1$ (see for comparison the monodromy arguments of \cite{DelDuca:2016lad}).

\vspace{\mylength}
%%%%%%%%%%%%%%%%%%%%%%%%%%%%%%%%%%%%%%%%%%%%%%%%%%%%%%%%%
%%%%%%%%%%%%%%%%%%%%%%%%%%%%%%%%%%%%%%%%%%%%%%%%%%%%%%%%%
\section{General multivariable result}
\label{sec:modten}
%%%%%%%%%%%%%%%%%%%%%%%%%%%%%%%%%%%%%%%%%%%%%%%%%%%%%%%%%
%%%%%%%%%%%%%%%%%%%%%%%%%%%%%%%%%%%%%%%%%%%%%%%%%%%%%%%%%
\vspace{\myotherlength}

This section gives our results for the motivic coaction and the single-valued map of MPLs in full generality, for any number $n$
of variables. The formulas are structurally analogous to those for
MPLs in two variables presented in the previous section. As we will see, the additional challenges of the $n$-variable case are mostly of combinatorial nature, in particular to present the explicit form of the braid algebra and its extension by zeta generators.

At the end of this section, we comment on how our general
results relate to the properties of period matrices, and explain how the new
results resonate with previous works on configuration-space integrals at genus zero subject to multivariable KZ equations.
 
 \medskip
\noindent
{\bf A. Notation:} We again specify a fibration basis for MPLs where the labels of $G(a_1,\ldots,a_r;z_n)$ are restricted to $a_i \in \{0,1,z_1,\ldots,z_{n-1} \}$. In order to state the general result, it is helpful to introduce the following shorthand for the generating series of $n$-variable MPLs
\begin{align}
\GG_n &:= \GG_{A=\{0,1,z_1,z_2,\ldots,z_{n-1}\}}(z_n;e_i^{(n)})
\label{nvars.01} \\
&\phantom{:}=  \sum_{r=0}^\infty \sum_{a_1,\ldots,a_r  \atop{\in \{0,1,z_1,z_2,\ldots,z_{n-1}\} } }
\! \! \! \! e^{(n)}_{a_1} \ldots e^{(n)}_{a_r} 
% G(a_1,\ldots, a_r;z_n) 
G(a_r,\ldots, a_1;z_n) 
\notag
\end{align}
In this notation $\GG_n$, the $n{+}1$ letters in the alphabet $A$
and the associated braid operators $e_i^{(n)}$ are left implicit.
In particular, for $n=1$ or $n=2$ variables, we now write 
\begin{align}
\GG_1 &= \GG_{\{0,1\}}(z_1;e_0^{(1)},e_{1}^{(1)}) \label{nvars.02} \\
\GG_2 &= \GG_{\{0,1,z_1\}}(z_2;e_0^{(2)},e_{1}^{(2)},e_{z_1}^{(2)})
\notag
\end{align}
with $e_i^{(1)}=e_i $ in earlier sections. In this notation, our formulas for $n=2$, (\ref{coact.42}) and (\ref{svcoact.42}), become
\begin{align}
\Delta \GG_2^\mot &=( \MM^{\dR} \GG_1^{\dR} )^{-1} \,\GG_2^\mot\, \MM^{\dR} \,\GG_1^{\dR} \GG_2^{\dR} 
\label{nvars.03} \\
 {\rm sv} \,  \GG_2 &=  {\rm sv} ( \MM \, \GG_1 )^{-1} \, \overline{ \GG_2^t}  \,{\rm sv} ( \MM \, \GG_1 )\, \GG_2  \notag
\end{align}

\medskip
\noindent
{\bf B. The results:} For MPLs in any number $n$ of variables, our main results for the motivic coaction and the single-valued map are generated by
\begin{align}
\Delta \GG_n^\mot &=( \MM^{\dR}\, \GG_1^{\dR} \ldots \GG_{n-1}^{\dR} )^{-1} \,\GG_n^\mot\, \MM^{\dR} \,\GG_1^{\dR}  \ldots \GG_{n-1}^{\dR}  \, \GG_n^{\dR} 
 \notag \\
 {\rm sv} \,  \GG_n &=  {\rm sv} ( \MM\,  \GG_1 \ldots \GG_{n-1} )^{-1} \, \overline{ \GG_n^t}  \,{\rm sv} ( \MM \, \GG_1 \ldots \GG_{n-1} )\, \GG_n 
 \label{nvars.04}
\end{align} 
where the second line recursively determines
$ {\rm sv} \,  \GG_n $ in terms
of generating series
$ {\rm sv} \,  \GG_{j} $ in fewer variables $j<n$.

As a key advantage of these formulas over earlier expressions for the coaction \cite{Goncharov:2001iea, Goncharov:2005sla, Brown:2011ik, BrownTate} and the single-valued map \cite{DelDuca:2016lad, Broedel:2016kls}, the right-hand sides are already in fully simplified form. First, the right-hand side of (\ref{nvars.04}) incorporates all known $\mathbb Q$-relations among MZVs by virtue of the $f$-alphabet representation (\ref{coact.15}) of $\mathbb M$. Second, all MPLs $G(b_1,\ldots,b_r;z_j)$ in $j\leq n$ variables on the right-hand side of (\ref{nvars.04}) are already in a uniform fibration basis with $b_i \in \{0,1,z_1,\ldots,z_{j-1} \}$.

The proof of our main results (\ref{nvars.04}) will be given in \cite{Frost:2024}, by demonstrating that the coaction formula is equivalent to the multivariate Ihara formula \cite{Brown:2019jng}, and that the ${\rm sv}$
formula is equivalent to the construction of single-valued polylogarithms in \cite{DelDuca:2016lad}.  
%\osnote{Although the individual zeta generators $M_{2k+1}$ and the $f$-alphabet used in intermediate steps of our construction depend on a choice of basis of MZVs, the proof in  \cite{Frost:2024} will expose that the final formulae for $\Delta G$ and $ {\rm sv} \, G$ do not depend on this basis choice.}
On the one hand, the individual zeta generators $M_{2k+1}$ and the $f$-alphabet used in intermediate steps of our construction depend on a choice of basis of MZVs, say (\ref{bMZVs}) in our case. On the other hand, the proof in \cite{Frost:2024} will expose that the expressions for $\Delta G$ and $ {\rm sv} \, G$ obtained from (\ref{nvars.04}) do not depend on this choice of basis.

In order to read off the coaction and single-valued map of individual $n$-variable MPLs, it remains to understand the braid algebra obeyed by the operators $e_{i}^{(j)}$ at $j \leq n$ and its extension by the zeta generators~$M_{2k+1}$. Using the commutation relations of this algebra, the series in (\ref{nvars.04}) can be systematically expanded in terms of solely the braid operators $e_i^{(n)}$ with $i \in \{0,1,z_1,\ldots,z_{n-1}\}$ at fixed $n$. Since the universal enveloping algebra of 
the $e^{(n)}_i$ entering $\GG_n$ in (\ref{nvars.01})
is freely generated, the coaction and single-valued map of $n$-variable MPLs $G(a_1,\ldots,a_r;z_n)$ is uniquely specified by the coefficients of $e_{a_r} \ldots e_{a_1}$ (\ref{nvars.04}).

Following the treatment of the two-variable case, we will present the commutation relations for the extended braid algebra in two steps in the next subsections: the Yang--Baxter relations for the braid algebras in the $n$-variable case in subsection {\bf C}, followed by the brackets of $e_i^{(n)}$ with zeta generators in subsection {\bf D}.

\medskip
\noindent
{\bf C. The braid algebra:} In the present notation, the braid algebra for the $n$-variables case is defined by the commutators $[e^{(n)}_a, e^{(m)}_b]$, with $m<n$. We group these commutators according to the following three cases:
\begin{itemize}
\item let $m<n$ and $j<n$, then:
\begin{align}
[e^{(n)}_0, e^{(m)}_0] &= 0\, ,
&[  e^{(n)}_1, e^{(m)}_0] &= [e^{(n)}_{1} ,e^{(n)}_{z_{m}}  ] \label{nvars.05} \\
[  e^{(n)}_0, e^{(m)}_1] &= 0 \, ,
&[  e^{(n)}_1, e^{(m)}_1] &= [e^{(n)}_{z_{m}} ,e^{(n)}_{1}  ]\notag \\
[  e^{(n)}_{0}, e^{(m)}_{z_j}] &= 0  \, ,
&[  e^{(n)}_1, e^{(m)}_{z_j}] &= 0
\notag
\end{align}
\item let $m<n$ and $i<n$, then:
\begin{align}
[e^{(n)}_{z_{i}}, e^{(m)}_0] &=
\begin{cases}
[e^{(n)}_{z_{i}}, e^{(n)}_{z_{m}}] \, , & i<m \\
\big[e^{(n)}_{0}+\sum_{j=m+1}^{n-1}e^{(n)}_{z_{j}}, e^{(n)}_{z_{m}}\big] \, , & i=m \\
0 \, , & i>m
\end{cases} \notag \\
[e^{(n)}_{z_{i}}, e^{(m)}_1] &=
\begin{cases}
[e^{(n)}_{1}, e^{(n)}_{z_{m}}] \, , & i=m \\
0 \, , & i \neq m
\end{cases} 
\label{nvars.07}
\end{align}
\item let $m<n$, $i<n$, and $j<m$, then:
\begin{align}
[e^{(n)}_{z_{i}}, e^{(m)}_{z_{j}}] &=
\begin{cases}
[e^{(n)}_{z_{m}}, e^{(n)}_{z_{i}}] \, , & i=j \\
[e^{(n)}_{z_{j}}, e^{(n)}_{z_{m}}] \, , & i=m \\
0 \, , & i\neq j,m
\end{cases} 
\label{nvars.08}
\end{align}
\end{itemize}
The key point is that these commutation relations reduce any commutator $[e^{(n)}_a, e^{(m)}_b]$ with $m<n$ to the braid operators $e^{(n)}_i$ that appear in the generating series $\GG_n$.

The commutation relations (\ref{nvars.05}) to (\ref{nvars.08}) follow from the Yang--Baxter relations (\ref{YBR}) under the identifications
\begin{align}
e_0^{(n)} &= x_{z_n 0} \, , \ \ \ \ e_1^{(n)} = x_{z_n 1}  \, , &\! \! \! \! \! e_{z_{j}}^{(n)} &= x_{z_n z_j}
\notag \\
e_0^{(m)} &= x_{z_m 0}+\sum_{j=m+1}^{n} x_{z_m z_j}  \, , &\! \! \! \! \! e_1^{(m)} &= x_{z_m 1} \, \notag \\
e_{z_{j}}^{(m)}&=x_{z_m z_j}  \, , \ \ \ \ m<n
\label{identYB-general}
\end{align}
to be motivated in the discussion around (\ref{PM.01}) below.

\medskip
\noindent
{\bf D. The extension by zeta generators:} To complete the description of the extended braid algebra for $n$ variables, we spell out the commutators of the zeta generators $M_{2k+1}$ with the braid operators $e^{(n)}_i$ entering $\GG_n$. The combinatorially more involved generalization of the one- and two-variable cases (\ref{coact.40}) and (\ref{coact.49}) is given by
\begin{align}
[e^{(n)}_{0} , M_{2k+1}] &= 0
\label{allMe}\\
[e^{(n)}_{z_{j}} , M_{2k+1}] &=  \Big[ W_{2k+1}(e_0^{(n)}+ \sum_{i=j+1}^{n-1} e_{z_i}^{(n)} , e^{(n)}_{z_{j}}) 
\notag\\
&\quad + \sum_{i=j+1}^{n-1} \Big\{ W_{2k+1}(e_0^{(n)}+ \sum_{r=i+1}^{n-1} e_{z_r}^{(n)} , e_{z_i}^{(n)})
\notag \\
&\quad \quad \quad \quad + W_{2k+1}(e_{z_i}^{(n)} , e^{(i)}_0 - e_{z_i}^{(n)})
\Big\} , e^{(n)}_{z_{j}} \Big]  \notag 
\end{align}
with $j=0,1,\ldots,n{-}1$, where we
define $e_{z_0}^{(n)}= e_{1}^{(n)}$
to also specify $[e^{(n)}_{1} , M_{2k+1}]$ through
the $j=0$ instance of the last three lines.
The sums over $i$ in the second and third line are absent when $j=n{-}1$. The Lie polynomials, $W_{2k+1}$, are defined by (\ref{defWk}). %
A derivation of the commutators ({\ref{allMe}) will appear in \cite{Frost:2024}, based on studying analytic continuations of the generating series for MPLs.

The key point for our present purposes is that the commutators ({\ref{allMe}), together with the Yang--Baxter relations, suffice to extract the motivic coaction and the single-valued map of individual MPLs from our main formulas in (\ref{nvars.04}). First, the commutators ({\ref{allMe}) can be applied in our formulas to eliminate the zeta generators $M_{2k+1}$ in favor of the braid operators $e_{i}^{(j)}$. The final step is to eliminate all braid operators $e_i^{(m)}$ with $m<n$ in favour of the $n{+}1$ generators $e_i^{(n)}$ that occur in $\GG_n$ by means of the braid algebra (\ref{nvars.05}) to (\ref{nvars.08}).

\medskip
\noindent
{\bf E. Relation to period matrices:} We briefly relate our results to earlier work through alternative presentations of our formulas. Consider the concatenation product
\begin{align}
\FFF_n = \MM \, \GG_1 \, \GG_2 \ldots \GG_n
\label{PM.01}
\end{align}
of the generating series $\GG_j$ at $1 \leq j \leq n$. This product $\FFF_n$ solves the multivariable KZ equations (\ref{altcoact.18}). The particular braid operators in (\ref{identYB-general}) can be identified by integrating this KZ connection along a specific path connecting the origin $(0,0,\ldots,0)$ and a generic point $(z_1,z_2,\ldots,z_n)$.\footnote{The required path is a series of straight lines moving each $z_i$ coordinate from $0$ to its non-zero value, in order from $i=1$ to $n$.} %: paramatrize $(t_1 z_1,\ldots,t_n z_n)$ and compose the
%straight paths 
%\begin{align}
%
%\! \! (t_1,t_2,\ldots,t_n) = \big( \underbrace{1,\ldots,1}_{j-1},t, \underbrace{0,\ldots,0}_{n-j} \big) \, , \ \ \ \ t \in (0,1)
%
%\end{align}
%in the order of $j=1,2,\ldots,n$ such that $z_1$ to $z_n$ are successively moved from
%zero to their actual values.
%\begin{align}
%%
%(0,\ldots,0,0)
%&\rightarrow
%(0,\ldots,0,z_1) 
%\rightarrow 
%(0,\ldots,0,z_2,z_1) 
%\rightarrow \ldots
%\\
%\nonumber
%&\rightarrow \ldots \rightarrow
%(0,z_{n-1},\ldots,z_1) \rightarrow
%(z_n,z_{n-1,}\ldots,z_1)
%%
%\end{align}}
 By induction on $n$, one can show that the main results (\ref{nvars.04}) of this work can be written compactly as
\begin{align}
\Delta \FFF_n^{\mot} &= \FFF_n^{\mot}  \, \FFF_n^{\dR}
 \, ,  \ \ \ \
{\rm sv} \, \FFF_n = \overline{\FFF_n^t} \, \FFF_n  \label{PM.02}
\end{align}
In practice, these attractive formulas are not as computationally useful as the formulas (\ref{nvars.04}) involving $\GG_n$. This is because $\FFF_n$ takes values in the universal enveloping algebra of all of $M_{2k+1}$ and $e_i^{(j)}$ with $j \leq n$ subject to a multitude of relations. To compare the coefficients of the words in ${e^{(j)}_{i}}$ and $M_{2k+1}$ appearing on both sides of the formulas in (\ref{PM.02})---in order to extract a formula for $\Delta G(a_1,a_2,\ldots,a_r;z_m)$ or ${\rm sv} \, G(a_1,a_2,\ldots,a_r;z_m)$---it is necessary to make exhaustive use of the extended braid-algebra commutators (\ref{nvars.05}) to (\ref{nvars.08}) and (\ref{allMe}) to line up the ordering of generators on the left- and right-hand sides. By contrast, the $n{+}1$ generators $e_i^{(n)}$ in $\GG_n$ generate a free algebra, and there is no need to specify any ordering convention when using our formulas, (\ref{nvars.04}), to compute $\Delta G(a_1,a_2,\ldots,a_r;z_m)$ or ${\rm sv} \, G(a_1,a_2,\ldots,a_r;z_m)$.

However, the formulas (\ref{PM.02}) help to illustrate the relation of our results to earlier work. In \cite{Britto:2021prf}, configuration-space integrals over marked points on a disk or sphere are shown to give rise to solutions to KZ equations with the same product structure as~$\FFF_n$ in (\ref{PM.01}). 
These solutions in the reference can be obtained from products $\mathbb P\,\FFF_n$ (with a generating series $\mathbb P =  1 + \sum_{k=1}^{\infty} \zeta_2^k P_{2k}$ in even zeta values)
by choosing particular matrix representations of the braid algebra, its extension by zeta generators $M_{2k+1}$, and the left-multiplicative matrices $P_{2k}$. In fact, our main results were initially anticipated from matrix representations of (\ref{PM.01}) investigated in \cite{Britto:2021prf}. However, in the present work, the Lie-algebra generators $e_i^{(j)}$ and $M_{2k+1}$ are studied without specialising to the matrix representations in the reference. This ensures that the generators satisfy only the extended braid-algebra relations (\ref{nvars.05}) to (\ref{nvars.08}) and (\ref{allMe}), without obeying any additional relations that arise as an artifact of the choice of representation.

The products $\mathbb P\,\FFF_n$ in \cite{Britto:2021prf} take the role of
twisted period matrices for configuration spaces at genus zero with different numbers of integrated and unintegrated marked points. 
By the corollary $\mathbb P^{\dR} = 1$ of $\zeta_{2k}^{\dR} = 0$, the period matrices $\mathbb P\,\FFF_n$ in the reference obey the same coaction properties as $\FFF_n$ in (\ref{PM.02}). Hence, the specializations of our main coaction formula in (\ref{PM.02}) to matrix representations yield examples of the so-called ``master formula'' \cite{Abreu:2017enx, Abreu:2017mtm} proposed for the motivic coaction of more general (twisted) period matrices, see \cite{Brown:2019jng, Kamlesh:2024} for proofs in certain cases.
Moreover, the formula (\ref{PM.02}) for ${\rm sv}\, \FFF_n$ lines up with the general theory of single-valued period matrices in \cite{Brown:2018omk}. When specializing to the matrix representations in \cite{Britto:2021prf},
%and taking the properties of $\mathbb P$ into account, 
the expression for ${\rm sv}\, \FFF_n$ in (\ref{PM.02}) can be derived from the generalization of the genus-zero KLT relations to an arbitrary number of unintegrated marked points \cite{Kawai:1985xq, Vanhove:2018elu}, with further details to be given in  \cite{Frost:2024}.

\vspace{\mylength}
%%%%%%%%%%%%%%%%%%%%%%%%%%%%%%%%%%%%%%%%%%%%%%%%%%%%%%%%%
%%%%%%%%%%%%%%%%%%%%%%%%%%%%%%%%%%%%%%%%%%%%%%%%%%%%%%%%%
\section{Conclusion and outlook}
%%%%%%%%%%%%%%%%%%%%%%%%%%%%%%%%%%%%%%%%%%%%%%%%%%%%%%%%%
%%%%%%%%%%%%%%%%%%%%%%%%%%%%%%%%%%%%%%%%%%%%%%%%%%%%%%%%%
\vspace{\myotherlength}

In this work, we have provided a new formulation of the motivic coaction and single-valued map for multiple polylogarithms (MPLs) in any number of variables. Our results are based on Lie-algebra structures that intertwine braid operators with additional generators for odd Riemann zeta values. The main result (\ref{nvars.04}) (which is specialised to the one- and two-variables cases in (\ref{coact.36}--\ref{coact.57}) and (\ref{coact.42}--\ref{svcoact.42})) is expressed in terms of generating series that combine these two types of generators with MPLs and MZVs. The coaction and single-valued map of the individual MPLs can be read off after applying the Lie-brackets of the generators that we determine explicitly from Yang--Baxter relations and the Drinfeld associator. Unlike earlier approaches, the MPLs and MZVs in our formulas are automatically fully simplified with respect to changes of fibration basis and MZV relations over $\mathbb Q$.

A longer companion paper \cite{Frost:2024} will give both heuristic derivations and proofs of the results reported here. In particular, the construction of the single-valued map from the coaction and the antipode \cite{Brown:2013gia, DelDuca:2016lad} will make manifest that all of these results are underpinned by the same extension of the braid algebra by zeta generators.

For the genus-zero MPLs studied in this work, the motivic coaction and the single-valued map already find a broad spectrum of applications to quantum-field-theory and string amplitudes. The coaction gives rise to smoking guns for new symmetries of physical theories under the so-called cosmic Galois group. 
The single-valued map in turn is a rich source of integration techniques and relations between open- and closed-string amplitudes.
However, both of these intriguing structures have mostly been studied at the level of genus-zero contributions to scattering amplitudes, i.e.\ MPLs and MZVs. Accordingly, investigations of coactions and single-valued maps of elliptic polylogarithms and more general functions beyond MPLs encountered in amplitudes will have important implications for understanding the symmetries of and relations among fundamental interactions.

Our reformulation of coactions and single-valued maps at genus zero via zeta generators is tailored to facilitate generalizations to elliptic polylogarithms and beyond. Indeed, Brown's equivariant and single-valued iterated Eisenstein integrals \cite{Brown:mmv, Brown:2017qwo, Brown:2017qwo2} at genus one give another class of functions where zeta generators control the appearance of MZVs in non-holomorphic modular forms. More specifically, Brown's construction can be rewritten in terms of adjoint actions of generating series similar to the one $ {\rm sv} \, \MM$ in this work \cite{Dorigoni:2022npe, Dorigoni:2023part2}, where the zeta generators act on certain derivations dual to Eisenstein series \cite{Ihara:1990, IharaTakao:1993, Tsunogai, Pollack}, instead of on the braid operators at genus zero. The close analogy between functions of the modular parameter at genus one and MPLs in one variable at genus zero will be highlighted in upcoming work \cite{Dorigoni:2023part2}, and this connection calls for future investigations of multivariable cases and implications for coaction formulas.

In the long run, our unified approach to the motivic coaction and single-valued map via Lie-algebra structures including zeta generators might offer a systematic line of attack for MPLs on surfaces of arbitrary genus. It appears promising to look for similar results using these methods for more general classes of iterated integrals depending on increasing numbers of marked points and complex-structure moduli.

\vspace{\mylength}
%%%%%%%%%%%%%%%%%%%%%%%%%%%%%%%%%%%%%%%%%%%%%%%%%%%%%%%%%
%%%%%%%%%%%%%%%%%%%%%%%%%%%%%%%%%%%%%%%%%%%%%%%%%%%%%%%%%
\section{Acknowledgements}
%%%%%%%%%%%%%%%%%%%%%%%%%%%%%%%%%%%%%%%%%%%%%%%%%%%%%%%%%
%%%%%%%%%%%%%%%%%%%%%%%%%%%%%%%%%%%%%%%%%%%%%%%%%%%%%%%%%
\vspace{\myotherlength}

We are grateful to Ruth Britto, Francis Brown, Daniele Dorigoni, Mehregan Doroudiani, Joshua Drewitt, James Drummond, Axel Kleinschmidt, Carlos Mafra, Lionel Mason, Sebastian Mizera, Pierre Lochak, Franziska Porkert, Leila Schneps, Oliver Schnetz and Yoann Sohnle for combinations of valuable discussions and collaboration on related topics.
The research of HF is supported by Merton College, Oxford, and received additional support from the European Research Council under ERC-COG-724638 GALOP.
The research of MH is supported in part by the European Research Council under ERC-STG-804286 UNISCAMP and in part by the Knut and Alice Wallenberg Foundation under grant KAW 2018.0116. 
The research of DK was supported by the European Research Council under ERC-COG-724638 GALOP.
The research of CR and OS is supported by the European Research Council under ERC-STG-804286 UNISCAMP.
The research of BV was supported by the Knut and Alice Wallenberg Foundation under grant KAW2018.0162.

\vskip -0.3cm

%\bibliographystyle{JHEP}
%\bibliography{refs}{}

\providecommand{\href}[2]{#2}\begingroup\raggedright\endgroup

\end{document}